\documentclass[12pt,epsf]{article}
\usepackage[pctex32]{graphics}
\makeatletter

\@addtoreset{equation}{section}
\newcommand{\be}{\begin{eqnarray}}
\newcommand{\ee}{\end{eqnarray}}
\newcommand{\ba}{\begin{array}}
\newcommand{\ea}{\end{array}}
\newcommand{\nn}{\nonumber}

\makeatletter \@addtoreset{equation}{section} \makeatother

\begin{document}
\vspace{1cm}
\begin{center}
~\\~\\~\\
{\bf  \LARGE Mass generation for Self-dual Gauge Fields via Topologically Coupling}
\vspace{1cm}

                      Wung-Hong Huang\\
                       Department of Physics\\
                       National Cheng Kung University\\
                       Tainan, Taiwan\\

\end{center}
\vspace{1cm}
\begin{center}{\bf  \Large ABSTRACT } \end{center}
We study the mass generation of self-dual two form gauge field by topologically coupling it to the three rank anti-symmetric field.  We use the various non-covariant Lagrangian of self-dual two form to perform the analysis and find that the field equation of massive self-dual two form gauge field may be described by $\partial^2 B^{ab}+m^2B^{ab}=0$ or $\partial_c (F^{cab}-\tilde F^{cab})+m^2B^{ab}=0$. We also find that, depending on the formulation of self-dual two form gauge field, the  three rank anti-symmetric field may or may not be totally eliminated.  Thus the different  Lagrangian will show different mass generation mechanisms.

\vspace{2cm}
\begin{flushleft}
*E-mail:  whhwung@mail.ncku.edu.tw\\
\end{flushleft}
\newpage
\section{Introduction}
In string theory [1] and M-theory five-branes [2] the Chiral p-forms, i.e. antisymmetric boson fields with self-dual (p+1)-form field strengths play a central role.  However, there is a problem in Lagrangian description of chiral bosons, since manifest duality and spacetime covariance do not like to live in harmony with each other in one action, as first seen by Marcus and Schwarz [3].

   Historically, the non-manifestly spacetime covariant action for self-dual  0-form was proposed by Floreanini and Jackiw [4], which is then generalized to p-form by Henneaux and Teitelboim [5].  Recently, a new non-covariant Lagrangian formulation of a chiral 2-form gauge field in 6D, called as (6=3+3) decomposition,  was derived in [6] from the Bagger-Lambert-Gustavsson (BLG) model [7].  Later, a general non-covariant Lagrangian formulation of  self-dual gauge theories in diverse dimensions was constructed [8].  In this general formulation the (6=2+4) decomposition of  Lagrangian was found. In a recent paper [9] we have found the new non-covariant Lagrangian formulation in the decomposition of  $D=D_1+D_2+D_3$ and other possible decomposition. 

In [10] Lambert et.al.  had discussed 6D Higgs mechanism of  self-dual two form and pointed some problems therein\footnote{We thank Kuo-Wei Huang for mentioning this problem to us}.  Let us first mention that manifest duality and mass do not like to live in harmony with each other.  The self-duality property of  the field strength implies following relation
\be \tilde H_{abc} =H_{abc}~~~&\Rightarrow&~~~{1\over 6}\epsilon_{abcdef}H^{def}=H_{abc}\nn\\
&\Rightarrow&~~~{1\over 2}\epsilon_{abcdef}\partial^d B^{ef}=\partial_a B_{bc}+\partial_c B_{ab}+\partial_b B_{ca}\nn\\
&\Rightarrow&~~~0={1\over 2}\partial^a (\epsilon_{abcdef}\partial^d B^{ef})=\partial^a(\partial_a B_{bc}+\partial_c B_{ab}+\partial_b B_{ca})\nn\\
&\Rightarrow&~~~0=\partial^a\partial_a B_{bc}+(\partial_c \partial^a B_{ab}+\partial_b \partial^a B_{ca})
\ee 
Under the Lorenz gauge $\partial^a B_{ab}=0$ above relation implies that $B_{bc}$ is described by massless wave equation.  We can arrive the same conclusion without use the Lorenz gauge.  In this case we define  mass of  $B_{bc}$ by 
\be \partial_c \partial^a B_{ab}+\partial_b \partial^a B_{ca}=m^2 B_{bc}~~~&\Rightarrow&~~~\partial_a (\partial_c \partial^s B_{sb}+\partial_b \partial^s B_{cs})=m^2 \partial_a  B_{bc}
\ee
In the same way, we get relations
\be \partial_b (\partial_a \partial^s B_{sc}+\partial_c \partial^s B_{as})=m^2 \partial_b  B_{ca}\\
\partial_c (\partial_b \partial^s B_{sa}+\partial_a \partial^s B_{bs})=m^2 \partial_c  B_{ab}\ee
Add above three relations we find that $0=m^2 H_{abc}$, which implies $m=0$ for self-dual two form gauge field.  Notice that in above proof we have assume that all components of $B_{ab}$ field have same mass.

In this paper we will discuss the mass generation of self-dual two form gauge field from topologically coupling method [11,12]. In this method we does not break the  gauge symmetry and is no need of a Higgs field.  We topologically couple self-dual two form gauge field to the three rank anti-symmetric field and use the various non-covariant Lagrangian of self-dual two form to analysis the mass generation mechanism therein.

In section 2 we first review the topologically massive BF model (TMBF) in which the Maxwell field strength $F$ is  topologically coupling to the two rank anti-symmetric field $B$ by the $\epsilon_{\mu\nu\lambda\rho}B^{\lambda\rho}F^{\mu\nu}$.  We see how the $B$ field is totally eliminated and photon acquires mass. Next, we extend the mechanism to the system with topologically couple self-dual two form gauge field to the three rank anti-symmetric field. 

In section 3 we first use the oldest non-covariant Lagrangian, the decomposition of  $6=1+5$ to discuss the issue of how to construct the topologically coupling of self-dual two form gauge field to the three rank anti-symmetric field.  In section 4 we use the decomposition of  $6=3+3$ in [6] to investigate the problem.  We have found that the field equation of massive self-dual two form gauge field may be described by $\partial^2 B^{ab}+m^2B^{ab}=0$ or
\be \partial_c (F^{cab}-\tilde F^{cab})+m^2B^{ab}=0\ee 
in which $\tilde F^{cab}$ is the dual field strength of $ F^{cab}$.   We also find that, depending on the formulation of self-dual two form gauge field, the  three rank anti-symmetric field may or may not be totally eliminated.  Thus different  Lagrangian will show different mass generation mechanism. Last section is devoted to a short conclusion and mention some furthermore works..  
\section {Topologically Coupling in 4D and 6D Spacetime}
We first review the 4D topologically massive BF model (TMBF) and then extend the model to higher rank field theory.

\subsection {Topologically Coupling in 4D Spacetime }
Action of topological massive BF model is
\be S=\int d^4x \Big[-{1\over4}F_{\mu\nu}F^{\mu\nu}+{1\over12}H_{\mu\nu\lambda}H^{\mu\nu\lambda}+{m\over 4}\epsilon^{\mu\nu\lambda\rho}B_{\mu\nu}F_{\lambda\rho}\Big]
\ee
where $F_{\mu\nu}\equiv \partial_{[\mu} A_{\nu]}$ and $H_{\mu\nu\lambda}\equiv \partial_{[\mu} B_{\nu\lambda]}$.  The action is invariant under the independent gauge transformations:
\be \delta A_\mu&=&\partial_\mu \Lambda,~~~\delta B_{\mu\nu}=0\\
\delta A_\mu&=&0 ,~~~~~~\delta B_{\mu\nu}=\partial_\mu \Lambda_\nu-\partial_\nu \Lambda_\mu
\ee
The field equations associated with  $A_\mu$ and $B_{\mu\nu}$ are
\be  \partial_\nu F^{\mu\nu}-{m\over6}\epsilon^{\mu\nu\lambda\rho}H_{\nu\lambda\rho}=0\\
\partial_\lambda H^{\mu\nu\lambda }+m\epsilon^{\mu\nu\lambda\rho}F_{\lambda\rho}=0
\ee
The solution of (2.5) is
\be   H^{\mu\nu\lambda }=\epsilon^{\mu\nu\lambda\rho}\Big(\partial_\rho S-m A_\rho\Big)
\ee
where $S$ is an arbitrary function.  Substituting the solution into (2.4) we find
\be  \partial_\nu F^{\mu\nu}-m^2 A^\mu +m~\partial^\mu S=0
\ee
Defining $\bar A_\mu\equiv A_\mu -{1\over m}\partial_\mu S$ above relation becomes
\be   \partial_\nu \bar F^{\mu\nu}-m^2 \bar A^\mu=0
\ee
and photon field $ A_\mu$ acquires mass.  Note that in 4D the massless anti-symmetric field $B^{\mu\nu}$ has one degree of freedom.  Now, through the  topologically coupling to the photon field by $\epsilon_{\mu\nu\lambda\rho}B^{\lambda\rho}F^{\mu\nu}$, the gauge invariant equations of motion (2.4) and (2.5) show that B field can be eliminated in terms of a gauge invariant vector field which satisfies the Proca equation and consequently we have only one massive spin-1 particle.\\

\subsection {Topologically Coupling in 6D Spacetime }
We can easily extend above mechanism to high rank tensor field.  Consider the action
\be S=\int d^6x \Big[-{1\over6}F_{\mu\nu\lambda}F^{\mu\nu\lambda}-{1\over48}W_{\mu\nu\lambda\rho}W^{\mu\nu\lambda\rho}+{m\over 3}\epsilon^{\mu\nu\lambda\rho\sigma\delta}W_{\mu\nu\lambda\rho}B_{\sigma\delta}\Big]
\ee
where $F_{\mu\nu\lambda}\equiv \partial_{[\mu} B_{\nu\lambda]}$ and $W_{\mu\nu\lambda\rho}\equiv \partial_{[\mu} C_{\nu\lambda\rho]}$.  The action is invariant under the independent gauge transformations:
\be \delta B_{\mu\nu}&=&\partial_{[\mu} \Lambda_{\nu]},~~~\delta C_{\mu\nu\lambda}=0\\
\delta B_{\mu\nu}&=&0 ,~~~~~~\delta C_{\mu\nu\lambda}=\partial_{[\mu} \Lambda_{\nu\lambda]}
\ee
The field equations associated with  $B_{\mu\nu}$ and $C_{\mu\nu\lambda}$ are
\be  \partial_\lambda F^{\mu\nu\lambda}-{m\over3}\epsilon^{\mu\nu\lambda\rho\sigma\delta}W_{\lambda\rho\sigma\delta}=0\\
\partial_\rho W^{\rho \mu\nu\lambda }-8m\epsilon^{\mu\nu\lambda\rho\sigma\delta} \partial_\rho B_{\sigma\delta}=0
\ee
The solution of (2.13) is
\be   W^{\mu\nu\lambda\rho}=\epsilon^{\mu\nu\lambda\rho\sigma\delta}\Big(8m B_{\sigma\delta}+\partial_{[\sigma} S_{\delta]})
\ee
where $S_\sigma$ is an arbitrary function.  Substituting the solution into (2.12) we find
\be  \partial_\lambda F^{\mu\nu\lambda}+{m\over3}\Big(16m~B^{\mu\nu}+\partial^{[\mu}S^{\nu]}\Big)=0
\ee
Defining $\bar B_{\mu\nu}\equiv B_{\mu\nu} -{1\over 16m}\partial^{[\mu}S^{\nu]}$ above relation becomes
\be   \partial_\lambda \bar F^{\mu\nu\lambda}-{16m\over3}~\bar B^{\mu\nu}=0
\ee
and anti-symmetry field $ B_{\mu\nu}$ acquires mass.  Note that in 6D the massless anti-symmetric field $C^{\mu\nu\lambda}$ has 4 degrees of freedom while massless anti-symmetric field $B_{\mu\nu}$ has 6 degrees of freedom.  Now, through the  topologically coupling by $\epsilon^{\mu\nu\lambda\rho ab}W_{\mu\nu\lambda\rho}B_{ab}$ we see that C field can be completely eliminated and consequently we have massive anti-symmetric field $B{\mu\nu}$, which has 10 degrees of freedom. 

In next section we will investigate how to extend above mechanism to the self-dual field theory

\section {Topologically Coupling in Self-Dual  2 form with 1+5 decomposition}
As there are many kinds of non-covariant Lagrangian for self-dual field [8,9] we will first consider the Lagrangian of (1+5) decomposition.  In this formulation the spacetime index $\mu= (1,\cdot\cdot\cdot,6)$ is decomposed as $(1,\dot a)$, with $\dot a=(2,\cdot\cdot\cdot,6)$. 
\subsection {Topologically Coupling  in 1+5 decomposition (A)}
The first possible action with self-dual field $B_{\mu\nu}$ which is topologically coupling to 3 rank antisymmetry field $C_{\mu\nu\lambda\rho}$ is described by the action
\be S_{1+5}=\int d^6x \Big[-{1\over4}\tilde F_{1\dot a\dot b} ( F^{1\dot a\dot b}-\tilde F^{1\dot a\dot b} ) -{1\over48}W_{\mu\nu\lambda\rho}W^{\mu\nu\lambda\rho}+{m\over 3}\epsilon^{\mu\nu\lambda\rho \sigma\delta}W_{\mu\nu\lambda\rho}B_{\sigma\delta}\Big]
\ee
in which the dual field strength is defined by$\tilde F^{\mu\nu\lambda} \equiv {1\over6}\epsilon^{\mu\nu\lambda\rho\sigma\delta} F_{\rho\sigma\delta}$. 

The variation of the action  with respect to field gives field equation.  First
\be 0={\delta S_{1+5}\over \delta B_{1\dot a}}&=&-{1\over2} \partial_{\dot b} \tilde F^{\dot b1 \dot a}+{2m\over3}\epsilon^{\dot b\dot c\dot d\dot e1\dot a}W_{\dot b\dot c\dot d\dot e}={2m\over3}\epsilon^{\dot b\dot c\dot d\dot e1\dot a}W_{\dot b\dot c\dot d\dot e}
\ee
in which we have used the property that $\partial_{\dot b} \tilde F^{\dot b1 \dot a}$ is identically zero.  Eq.(3.2) gives us a constrain 
\be
W_{\dot b\dot c\dot d\dot e}=0
\ee
This means that terms in action involved $B_{1\dot a}$ only through total derivative terms after using above constrain  and we have gauge symmetry
\be   \delta B_{1\dot a}=\Phi_{1\dot a}
\ee
for arbitrary functions $\Phi_{1\dot a}$. The gauge symmetry is crucial to prove the self-dual property in [8,9].  

The next field equation is
\be
0={\delta S_{1+5}\over \delta B_{\dot a\dot b}}&=&{1\over4}\Big(\partial_{1} \tilde F^{1\dot a \dot b}-\partial_{\dot c} \tilde F^{\dot c\dot a \dot b}\Big)+{1\over2} \partial_{\dot c} F^{\dot c\dot a \dot b}+{4m\over3}\epsilon^{1\dot a\dot b\dot c\dot d \dot e}W_{1\dot c \dot d\dot e}\nn\\
&=&{1\over2} \partial_{\dot c} {\cal F}^{1\dot a\dot b}+{4m\over3}\epsilon^{1\dot a\dot b\dot c\dot d \dot e}W_{1\dot d\dot e}
\ee
in which we define
\be  {\cal F}_{\mu\nu\lambda} \equiv F_{\mu\nu\lambda}-\tilde F_{\mu\nu\lambda}
\ee 
and use the property that $\partial_{1} \tilde F^{1\dot a \dot b}+\partial_{\dot c} \tilde F^{\dot c\dot a \dot b}=0$.  

The remained field equations are
\be
0={\delta S_{1+5}\over \delta C_{1\dot a\dot b}}&=&{1\over2} \partial_{\dot c} W^{\dot c1\dot a \dot b}-4m~\epsilon^{1\dot a\dot b\dot c\dot d \dot e}\partial_{\dot c}B_{\dot d\dot e}\\
0={\delta S_{1+5}\over \delta C_{\dot a\dot b\dot c}}&=&{1\over6} \partial_{1} W^{1\dot a\dot b b\dot c}-{4m\over3}\epsilon^{1\dot a\dot b\dot c\dot d \dot e}\partial_{1}B_{\dot d\dot e}
\ee
To obtain the last relation we have used (3.3) and (3.4). Eq.(3.7) has solution
\be W^{1\dot a\dot b \dot c}= \epsilon^{1\dot a\dot b\dot c\dot d \dot e}\Big(\partial_{\dot d}S_{\dot e}+8m B_{\dot d\dot e}\Big)
\ee
where $S_{\dot a}$ is an arbitrary function.  Note that above solution also solves (3.7).

 Finally, substituting the solution (3.9) into (3.5) we find
\be  \partial_{\dot c} {\cal F}^{\dot c\dot a\dot b}+{8m\over3}\Big(8m B^{\dot a\dot b}+\partial^{[\dot a}S^{\dot b]}\Big)=0
\ee
Define $\bar B^{\dot a\dot b}=B^{\dot a\dot b}+{1\over 8m}\partial^{[\dot a}S^{\dot b]}$ above equation becomes
\be  \partial_{\dot c} {\bar{\cal F}}^{\dot c\dot a\dot b}+{64m^2\over3}\bar B^{\dot a\dot b}=0
\ee
This describes massive the self-dual field which acquires mass through the topologically coupling with 3 form field.\\

Let us make following comments to conclude this subsection.

1. The massive self-dual field equation is $ \partial_{\dot c} \Big( F^{\dot c\dot a\dot b}- \tilde F^{\dot c\dot a\dot b}\Big)+{m^2}\bar B^{\dot a\dot b}=0$ while field equation of 2 form field without self-duality is $\partial_\lambda F^{\mu\nu\lambda}-{m^2}~B^{\mu\nu}=0$. Thus the field strength $F^{\dot c\dot a\dot b}$ in non-self-duality 2 form is replaced by the anti-self-dual field strength ${\cal F}^{\dot c\dot a\dot b}\equiv F^{\dot c\dot a\dot b}-\bar F^{\dot c\dot a\dot b}$ in massive self-dual 2 form field equation.

2. In 6D the massless self-dual 2 form has 3 degrees of freedom while massless anti-symmetric field $C_{\mu\nu}$ has 4 degrees of freedom.  However, under the constrain relation (3.3) $C_{\mu\nu}$ remain  3 degrees of freedom.  This means that we remain only 3+3=6 degrees of freedom.  Now, through the  topologically coupling we see that $B_{1\dot a}$ can be gauge alway and we remain a massive field $\bar B_{\dot a\dot b}$, which has 6 degrees of freedom.  Therefore, the degrees of freedom is conserved under topologically coupling.

3. The Lagrangian we used in (3.1) has the property that the kinetic part of self-dual 2 form is non-covariant while 3 form and  topologically coupling are covariant.  At first sight, this seems to be a nice choice as we have most terms which are covariant.  However, this choice suffers a strange property that the constrain relation (3.3) will delete one degrees of freedom from the original Lagrangian.  This lead us to consider another possible Lagrangian in which 3 form remains covariant but topologically coupling is non-covariant.  Note that as the kinetic part of self-dual 2 form is non-covariant the topologically coupling which couples 3 form with 2 form may be non-covariant. We will use this kind of  topologically coupling in the following section.

\subsection {Topologically Coupling  in 1+5 decomposition (B)}
We consider another possible action with self-dual field $B_{\mu\nu}$ which is topologically coupling to 3 rank antisymmetry field $C_{\mu\nu\lambda\rho}$. The action is  
\be S_{1+5}=\int d^6x \Big[-{1\over4}\tilde F_{1\dot a\dot b} ( F^{1\dot a\dot b}-\tilde F^{1\dot a\dot b} ) -{1\over48}W_{\mu\nu\lambda\rho}W^{\mu\nu\lambda\rho}+{m\over 3}\epsilon^{1\dot a\dot b\dot c\dot d\dot e}W_{1\dot a\dot b\dot c}B_{\dot d\dot e}\Big]
\ee
in which both of kinetic part of self-dual 2 form and topologically coupling of 3- form to 2-form are non-covariant.

The variation of the action  with respect to field gives field equation.  First
\be 0={\delta S_{1+5}\over \delta B_{1\dot a}}&=&-{1\over2} \partial_{\dot b} \tilde F^{\dot b1 \dot a}
\ee
which is identically zero. This means that terms involved $B_{1\dot a}$ only through total derivative terms and we have a  gauge symmetry
\be   \delta B_{1\dot a}=\Phi_{1\dot a}\ee
for arbitrary functions $\Phi_{1\dot a}$. The other field equations are
\be
0={\delta S_{1+5}\over \delta B_{\dot a\dot b}}&=&{1\over2} \partial_{\dot c} \Big(F^{\dot c\dot a \dot b}-\tilde F^{\dot c\dot a \dot b}\Big)+{m\over3}\epsilon^{1\dot a\dot b\dot c\dot d \dot e}W_{1\dot c\dot d\dot e}\\
0={\delta S_{1+5}\over \delta C_{1\dot a\dot b}}&=&{1\over2} \partial_{\dot c} W^{\dot c1\dot a \dot b}-m\epsilon^{1\dot a\dot b\dot c\dot d \dot e}\partial_{\dot c}B_{\dot d\dot e}\\
0={\delta S_{1+5}\over \delta C_{\dot a\dot b\dot c}}&=&{1\over6} \partial_{\dot d} W^{\dot d\dot a \dot b\dot c}+{1\over6} \partial_{1} W^{1\dot a\dot b \dot c}-{m\over3}\epsilon^{1\dot a\dot b\dot c\dot d \dot e}\partial_{1}B_{\dot d\dot e}
\ee
 Eq.(3.16) has solution
\be W^{1\dot a\dot b \dot c}= \epsilon^{1\dot a\dot b\dot c\dot d \dot e}\Big(\partial_{\dot d}S_{\dot e}+2m B_{\dot d\dot e}\Big)
\ee
where $S_{\dot a}$ is an arbitrary function. 

Substituting the solution (3.18) into (3.17) we find that
\be  \partial_{\dot d} W^{\dot d\dot a \dot b\dot c}+\epsilon^{1\dot a\dot b\dot c\dot d \dot e}\partial_{1}\partial_{\dot d}S_{\dot e}=0
\ee
which has solution
\be  W^{\dot d\dot a \dot b\dot c} = \epsilon^{\dot a\dot b\dot c\dot d \dot e}\Big(\partial_{\dot e}\Phi+S_{\dot e} \Big)
\ee
where $\Phi$ is an arbitrary function. This solution tells us that there is one degree freedom in non-zero value of $W^{\dot d\dot a \dot b\dot c}$.

Substituting the solution (3.18) into (3.15) we find that
\be  \partial_{\dot c} {\cal F}^{\dot c\dot a\dot b}+{2m\over3}\Big(16m B^{\dot a\dot b}+\partial^{[\dot a}S^{\dot b]}\Big)=0
\ee
Define $\bar B^{\dot a\dot b}=B^{\dot a\dot b}+{1\over 16m}\partial^{[\dot a}S^{\dot b]}$ above equation becomes
\be  \partial_{\dot c} {\bar{\cal F}}^{\dot c\dot a\dot b}+{32m^2\over3}\bar B^{\dot a\dot b}=0
\ee
This describes a massive self-dual field which acquires mass through the topologically coupling with 3 form field.\\

Note that in 6D the massless self-dual 2 form has 3 degrees of freedom while massless anti-symmetric field $C_{\mu\nu}$ has 4 degrees of freedom.  Under the topologically coupling there remains 1 degrees of freedom in $W_{\dot a\dot b\dot c\dot d}$ and we have a massive field $\bar B_{\dot a\dot b}$ which has 6 degrees of freedom.  Therefore, the degrees of freedom is conserved under topologically coupling.

\section {Topologically Coupling in Self-Dual 2 form with  3+3 decompositions}
In the (3+3) decomposition [6] the spacetime index $A$ is decomposed as $\mu=(a,\dot a)$, with $a=(1,2,3)$ and $\dot a=(4,5,6)$.  The action  we adopted is 
\be S_{3+3} &=&  \int dx^6\Big[-{1\over12}\tilde F_{abc} {\cal F}^{abc}-{1\over4}\tilde F_{ab\dot a} {\cal F}^{ab\dot a}-{1\over48}W_{\mu\nu\lambda\rho}W^{\mu\nu\lambda\rho}\nn\\
&&+m~\epsilon^{abc\dot a\dot b\dot c}\Big(2W_{bc\dot b\dot c}B_{a\dot a}+W_{abc\dot c}B_{\dot a\dot b}\Big)\Big]
\ee
The variation of the action  with respect to field gives field equation.  First
\be 0={\delta S_{3+3}\over \delta B_{ab}}&=&{1\over4} \Big(\partial_{c} \tilde F^{cab}+\partial_{\dot a} \tilde F^{ \dot aab}\Big)
\ee
which is identically zero and we have a  gauge symmetry
\be   \delta B_{ab}=\Phi_{ab}\ee
for arbitrary functions $\Phi_{ab}$. The other field equations are
\be
0={\delta S_{3+3}\over \delta B_{a\dot a}}&=&\partial_{\dot b} {\cal F}^{a \dot a\dot b}+2m~\epsilon^{abc\dot a\dot b\dot c}W_{bc\dot b\dot c}\\
0={\delta S_{3+3}\over \delta B_{\dot a\dot b}}&=&{1\over2} \Big(\partial_{\dot c} F^{\dot a\dot b\dot c}+\partial_a F^{a\dot a\dot b}\Big)+m~\epsilon^{abc\dot a\dot b\dot c}W_{abc\dot c}
\ee
The field equation
\be
0={\delta S_{3+3}\over \delta C_{abc}}&=&{1\over6} \partial_{\dot a} W^{\dot a abc}-m\epsilon^{abc\dot a\dot b\dot c}\partial_{\dot c}B_{\dot a\dot b}
\ee
has solution 
\be  W^{\dot a abc}= \epsilon^{abc\dot a\dot b\dot c}\Big(\partial_{\dot b}\Phi_{\dot c} +6m B_{\dot b\dot c}\Big)
\ee

Substituting the solution into (4.5) we find
\be 0={\delta S_{3+3}\over \delta B_{\dot a\dot b}}&=&{1\over2} \Big(\partial_{\dot c} F^{\dot a\dot b\dot c}+\partial_a F^{a\dot a\dot b}\Big)+4m~\Big(\partial^{[\dot a}\Phi^{\dot b]} +6m B^{\dot a\dot b}\Big)  
\ee
After proper gauge transition $\bar B_{\dot a\dot b}= B_{\dot a\dot b}+{1\over6m}\partial_{[\dot a} \Phi_{\dot b]}$ above relation becomes
\be  0=\partial_{\dot c} F^{\dot a\dot b\dot c}+\partial_a F^{a\dot a\dot b}+24m^2 \bar B^{\dot a\dot b}
\ee
which describes a massive field of 2 form field $\bar B^{a\dot b}$. In the Lorentz gauge above equation becomes 
\be \partial^2 B^{\dot a\dot b}+ 24m^2B^{\dot a\dot b}=0
\ee
which describes a massive wave equation.

The remained field equations are  
\be
0={\delta S_{3+3}\over \delta C_{\dot a\dot b\dot c}}&=&{1\over6} \partial_{a } W^{a\dot a \dot b\dot c}\\
0={\delta S_{3+3}\over \delta C_{ab\dot a}}&=&{1\over2} \partial_{c} W^{cab\dot a}+{2} \partial_{\dot b} W^{\dot b ab\dot a}-4m\epsilon^{abc\dot a\dot b\dot c}\partial_{\dot c}B_{c\dot b}-3m\epsilon^{abc\dot a\dot b\dot c}\partial_{c}B_{\dot b\dot c}\\
0={\delta S_{3+3}\over \delta C_{a\dot a\dot b}}&=&{4} \partial_{b} W^{ba\dot a \dot b}+{2} \partial_{\dot c} W^{\dot c a\dot a\dot b}-4m\epsilon^{abc\dot a\dot b\dot c}\partial_{b}B_{c\dot c}
\ee
in which $\Phi$ is an arbitrary function.  Substituting the solution (4.7) into (4.12) we can find solution of $W^{ab\dot a\dot c}$ 
\be  W^{ab\dot a\dot c}= \epsilon^{abc\dot a\dot b\dot c}\Big(\partial_{\dot b}\tilde \Phi_c+ 4m B_{c\dot b}-\partial_{\dot b}\Phi_c \Big)
\ee
in which $\tilde \Phi$ is another arbitrary function.

 Substituting the solution (4.14) into (4.4) we find 
\be 0=\partial_{\dot b} {\cal F}^{a \dot a\dot b}+2m \Big(4m B_{a\dot b}+\partial_{[a}\tilde \Phi_{\dot b]}-\partial_{[a}\Phi_{\dot b]} \Big)~\
\ee
After proper gauge transition $\bar B_{a\dot b}= B_{a\dot b}+{1\over4m}\Big(\partial_{[a}\tilde \Phi_{\dot b]}-\partial_{[a}\Phi_{\dot b]}\Big)$ above relation becomes
\be  0=\partial_{\dot b} {\bar {\cal F}}^{a \dot a\dot b}+8m^2 \bar B_{a\dot b}
\ee
which describes a massive field of 2 form field $\bar B_{a\dot b}$.

  Finally, Substituting the solution (4.14) into (4.13) we find that 
\be  0=\partial_{\dot c} W^{a\dot a\dot b \dot c}
\ee
Eqs.(4.11) and (4.17) describe field equation of  massless  $C_{\dot a\dot b\dot c}$ and  $C_{a\dot a\dot b}$ which, however, have no degree of freedom.

We thus  see that in 3+3 decomposition, the massless 3-rank field  $C_{abc}$ has 1 degree of freedom and $C_{ab\dot a}$ has 3 degree of freedoms. They are all absorbed by self-dual 2 form.  Now, through topologically coupling we find a massive $B_{a\dot a}$ which has 6 degree of freedoms and a massive $B_{\dot a\dot b}$ which has 1 degree of freedoms.   Therefore, the degrees of freedom is conserved under topologically coupling. Notice that in 1+5 decomposition we remain a massless $C$ contain which has 1 degree of freedoms after the topologically coupling,  while in 3+3 decomposition all degree of freedoms of $C$ field are eliminated. Thus, we see that,depending on the formulation of self-dual two form gauge field, the  three rank anti-symmetric field may or may not be totally eliminated and the different  Lagrangian will show different mass generation mechanism.\\ 

Let us mention some properties found from above investigation. 

1.  Without topologically coupling some components of 2 form in non-covariant Lagrangian only appear as a totally derivative terms. (for example the $B_{ab}$ field in (3.14)) When we consider topologically coupling of 2 form to 3 form the topological term shall not contain this $B_{ab}$ field. Otherwise, it will produce a strange constrain, like (3.3), and it will remove some degree of freedom in the original Lagrangian.

2. Without topologically coupling some field equations of self-dual  2 form may appear in a simple form : $0={\delta S\over \delta B^{ab}}=\partial_c {\cal F}^{cab}$. In this case, after the topologically coupling  the field equation of the massive self-dual 2 form will become $0=\partial_c {\cal F}^{cab}+M^2B^{ab}$, as those in (3.22) and (4.16).  Otherwise the field equation of the massive self-dual 2 form will become a massive wave equation, as that in (4.10).

3. It is expected that above properties will also show in other decomposition of self-dual  2 form and higher-rank tensor theory.

\section {Conclusion}  
In conclusion, we have discussed the mass generation of self-dual two form gauge field by considering it topologically couples to the three rank anti-symmetric field.  We have used 1+5 and 3+3 decompositions in the non-covariant Lagrangian of self-dual two form to analysis mass generation mechanism therein.  

We have  found that the field equation of massive self-dual two form gauge field may be described by $\partial^2 B^{ab}+m^2B^{ab}=0$ or $\partial_c (F^{cab}-\tilde F^{cab})+m^2B^{ab}=0$.  We also find that, depending on the formulation of self-dual two form gauge field, the  three rank anti-symmetric field may or may not be totally eliminated.  Thus, the different  Lagrangian will show different mass generation mechanism.  

Note that,  as there are many different formulations of the self-dual gauge field [8,9]  it will be interesting to investigate how the different  formulations will affect the mass generation mechanism in detail. It is also interesting to see how the mass generation mechanism adopted in this paper could be extended to be a covariant form, as that in PST formulation [13,14]. 

   Finally, in recent the non-Abelian self-dual gauge theory in 1+5 dimension had been found [15], i.e. HHM formulation, in which the  covariant derivative $D_\mu$ of self-dual two form is obtained therefore. Maybe one can use the covariant derivative to couple self-dual two form to scalar field and see how the Higgs mechanism will be shown in the HHM formulation.  It is interesting to compare this result with our finding in this paper.
\\
\\
\begin{center} {\bf REFERENCES}\end{center}
\begin{enumerate}
\item M. Green, J. Schwarz and E. Witten,``Superstring theory'', Cambridge University Press, Cambridge, 1987.
\item C.G. Callan, J.A. Harvey, and A. Strominger, Nucl. Phys. B367, 60 (1991);\\ E. Witten,`` Five brane effective action'', hep-th/9610234 [hep-th/9610234].
\item N. Marcus and J.H. Schwarz, Phys. Lett. 115B (1982) 111;\\J. H. Schwarz and A. Sen, ``Duality symmetric actions," Nucl. Phys. B 411, 35 (1994) [arXiv:hep-th/9304154].
\item R. Floreanini and R. Jackiw,``Selfdual fields as charge density solitons,'' Phys. Rev. Lett. 59 (1987) 1873.
\item M. Henneaux and C. Teitelboim, ``Dynamics of chiral (selfdual) p forms,''  Phys. Lett. B 206 (1988) 650.
\item P. M. Ho, Y. Matsuo, ``M5 from M2'', JHEP 0806 (2008) 105 [arXiv: 0804.3629 [hep-th]];  \\P.M. Ho, Y. Imamura, Y. Matsuo, S. Shiba, ``M5-brane in three-form flux and multiple M2-branes,'' JHEP 0808 (2008) 014, [arXiv:0805.2898 [hep-th]].
\item  J. Bagger and N. Lambert, ``Modeling multiple M2'' Phys. Rev. D 75 (2007) 045020 [arXiv:hep-th/0611108];\\
J.A. Bagger and N. Lambert, ``Gauge Symmetry and Supersymmetry of Multiple M2-Branes,'' Phys. Rev. D 77 (2008) 065008 arXiv:0711.0955 [hep-th];\\
J. Bagger and N. Lambert, ``Comments On Multiple M2-branes,'' JHEP 0802 (2008) 105, arXiv:0712.3738 [hep-th]; \\A. Gustavsson, ''Algebraic structures on parallel M2-branes,'' Nucl. Phys. B 811 (2009) 66, arXiv:0709.1260 [hep-th].
\item W.-M. Chen and P.-M. Ho,``Lagrangian Formulations of Self-dual Gauge Theories in Diverse Dimensions,'' Nucl. Phys. B837 (2010) 1 [arXiv:1001.3608 [hep-th]]. 
\item Wung-Hong Huang,``Lagrangian of Self-dual Gauge Fields in Various Formulations,'' Nucl. Phys. B861 (2012) 403 [arXiv:1111.5118 [hep-th]]. 
\item  N. Lambert, C. Papageorgakis, M. Schmidt-Sommerfeld,``M5-Branes, D4-Branes and Quantum 5D super-Yang-Mills,'' JHEP 1101 (2011) 083 [arXiv:1012.2882 [hep-th]]. 
\item T. J. Allen , M. J. Bowick and A. Lahiri,``Topological Mass Generation in 3+1 Dimensions,'' Mod. Phys. Lett. A6 (1991) 559; Phys. Lett. B237 (1989) 47.
\item M. Henneaux, V. E. Lemes, C. A. Sasaki, S. P. Sorella, O. S. Ventura and L. C. Vilar,``A No-Go Theorem for the Nonabelian Topological Mass Mechanism in Four Dimensions,'' Phys Lett. B410 195 (1997) [hep-th/9707129]; \\D. S. Hwang and C. Lee, ``Nonabelian Topological Mass Generation in 4 Dimensions, '' J.Math.Phys. 38 (1997) 30-38 [hep-th/9512216].
\item P. Pasti, D. Sorokin and M. Tonin,``Note on manifest Lorentz and general coordinate invariance in duality symmetric models,''  Phys. Lett. B 352 (1995) 59
[arXiv:hep-th/9503182];
P. Pasti, D. Sorokin and M. Tonin,``Duality symmetric actions with manifest
space-time symmetries, '' Phys. Rev. D 52 (1995) R4277 [arXiv:hep-th/9506109];
P. Pasti, D. Sorokin and M. Tonin,``Space-time symmetries in duality symmetric
models,'' [arXiv:hep-th/9509052]; P. Pasti, D. P. Sorokin, M. Tonin,``On Lorentz invariant actions for chiral p-forms,`` Phys. Rev. D 55 (1997) 6292 [arXiv:hepth/ 9611100].
\item P. Pasti, I. Samsonov, D. Sorokin and M. Tonin,``BLG-motivated Lagrangian formulation for the chiral two-form gauge field in D=6 and M5-branes, '' Phys. Rev. D 80, 086008 (2009) [arXiv:0907.4596 [hep-th]].
\item P.-M. Ho, K.-W. Huang and Y. Matsuo,`` A Non-Abelian Self-Dual Gauge Theory in 5+1 Dimensions, ''JHEP 1107 (2011) 021 [arXiv:1104.4040 [hep-th]].
\end{enumerate}
\end{document}